\documentclass{aa501}
\usepackage{graphicx}

\begin{document}

\title{Multiwavelength evidence for a 15-year periodic activity in
the symbiotic nova V1016 Cygni
\thanks{Partly based on observations obtained with the
International Ultraviolet Explorer (IUE) satellite retrieved from
the IUE Newly Extracted Spectra (INES) Archive and the Hopkins
Ultraviolet Telescope (HUT) retrieved from the Multimission Archive
at STScI (MAST).}}

\author{\v{S}.~Parimucha\inst{1,2} \and D.~Chochol\inst{2} \and
T.~Pribulla\inst{2} \and L.M.~Buson\inst{3} \and A.A.~Vittone\inst{3}}

\offprints{\v{S}.~Parimucha}

\institute{Faculty of Natural Sciences, Department of the Theoretical
Physics and Geophysics, University of P.J. \v{S}af\'arik, 040 01
Ko\v{s}ice, Slovakia, e-mail: parimuch@ta3.sk \and
Astronomical Institute of the Slovak Academy of Sciences, 059 60
Tatransk\'a Lomnica, Slovakia \and INAF Osservatorio Astronomico
di Capodimonte, Via Moiariello 16, 80131 Napoli, Italy}

\date{Received.... ; Accepted....}

\abstract{ The $\sim$15.1 years period found in the long-term $UBV$
photoelectric and photographic photometry of the symbiotic nova V1016 Cyg is
detected also in the $(J-K)$ colour index and in the UV continuum and emission
line fluxes from IUE and HUT spectra. It could be interpreted either as the
effect of recurrent enhanced mass loss episodes from the Mira type variable
companion to a hot component along its ultra-wide orbit (proposed from recent
HST observations) or the true orbital period of the inner, unresolved binary of
a triple system. A 410-day delay of the maximum of UV emission lines fluxes
with respect to the maximum of continuum was found. The pulsation period of the
Mira type variable was improved to 474$\pm$6 days.
 }

\authorrunning{\v{S}.~Parimucha et al.}

\titlerunning{15-year Periodic Activity in V1016 Cygni}

\maketitle

\section{Introduction}

V1016 Cyg (MH$\alpha$ 328-116) is a member of a small subgroup of symbiotic
stars, called symbiotic novae, also including V1329 Cyg and HM Sge, whose
outbursts lead to a nebular spectrum (M\"urset \& Nussbaumer \cite{mur}).
Symbiotic novae are wide interacting binaries, where matter from a late-type
giant is transferred onto the surface of the more compact companion. The
nova-like optical outburst ($\Delta m \sim 5-7$ mag), lasting decades, is
caused by a thermonuclear runaway on the surface of a wind-accreting white
dwarf after the critical amount of material has been accumulated (cf.
Mikolajewska \& Kenyon, \cite{mik}). V1016 Cyg underwent such nova-like
outburst in~1964 (McCuskey \cite{mcc}). The object is classified as a D-type
symbiotic, the cool component being a Mira type variable embedded in a dust
envelope whose pulsation period turned out to be $\sim$478 (Munari \cite{mun}).
The onset of a dust formation episode in 1983 is reported by Taranova \& Yudin
(\cite{tar2}).

The orbital period of V1016 Cyg is not yet established. Taranova \& Yudin
(\cite{tar1}) made use of the increase of Balmer emission lines in combination
with the appearance and disappearance of FeII lines to derive an orbital period
of $\approx$20 years. Afterwards, Nussbaumer \& Schmid (\cite{nuss2}), though
unable of recording two consecutive maxima, proposed an orbital period of 9.5
years on the basis of the apparent periodicity seen in the flux of OI and MgII
UV emission lines by means of the IUE satellite. At the same time Munari
(\cite{mun}), by resorting to IR observations taken over two decades, proposed
instead a 6-year orbital period by modeling the sequence of dust obscuration
episodes, likely related to the passage of the Mira type variable at the
inferior conjunction in the system.

Much longer periods have been proposed by Wallerstein (\cite{wall}) and Schild
\& Schmid (\cite{schi}). In the former paper the author, assuming that the sharp
FeII emission lines are formed in the chromosphere of the cool star so as to
reflect its orbital motion, concludes that their observed radial velocities
between 1978 and 1985 limit any high inclination orbit to a period greater than
25 years or to a large eccentricity. The latter analysis, based on spectropolarimetric
data taken from 1991 to 1994, indicates that the orbital period is about 80 $\pm$
25 years, though later observations obtained in 1997 put this result into
question (Schmid, \cite{schm}). Finally, Brocksopp et al. (\cite{bbe}), adopting a
projected separation of the two stellar components as large as 84 AU
on the basis of their HST/WFPC2 images, are forced to propose
an astonishingly long orbital period of $\sim$544 years.

\begin{figure*}
\centerline{\resizebox{18cm}{!}{\includegraphics{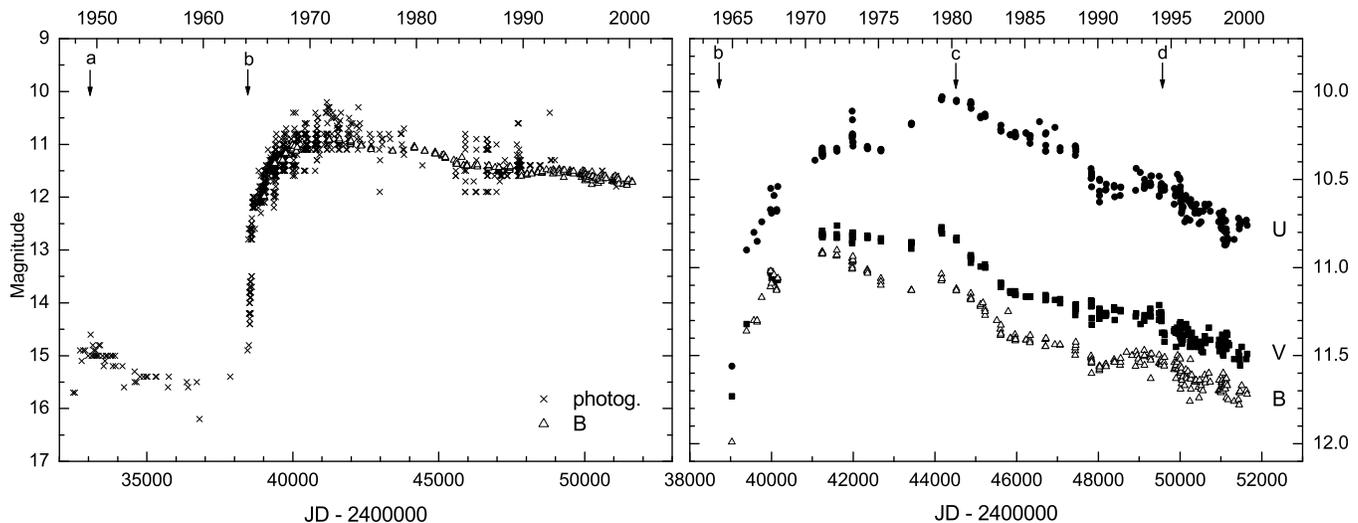}}}
\vspace*{-4.5cm}
\hfill \caption{Photographic (left) and $UBV$ (right) light curves
of V1016 Cyg. Arrows $a,b,c$ and $d$ mark the epochs of subsequent
activity episodes of the system (see text).}
\end{figure*}

In the context of this still-to-be-settled issue, one has also to interpret the
evidence for the 15-year periodic activity (not necessarily representing the
sought {\em orbital} period) recently presented by Parimucha et al. (\cite{pac})
(hereafter PAC) who gathered and analysed long-term photographic, photoelectric
and visual photometry of the object. The aim of the present paper is indeed to
give further (i.e., multiwavelength) support to the existence of such a periodicity
making use of both IR photometry and UV spectroscopy, as well as to investigate
its own origin.

\section{Long-term photometry}

\subsection{$UBV$ photoelectric and photographic data}

The historical light curve of V1016 Cyg based on the photographic and $UBV$
data given in PAC is presented in Fig.~1. The light curve suggests four
stages of activity marked by arrows in the figure: the pre-outburst flare
$a$ in 1949, the main nova-like outburst $b$ in 1964 and two post-outburst,
decreasing-amplitude flares $c$ and $d$ in 1980 and 1994, respectively.
Evidently the activity episodes affecting the system repeat themselves
(though at quite a different intensity level) with an interval of
$\approx$15 years. The ephemeris for the activity maxima calculated in PAC
is as follows:

\begin{center}
\begin{equation}
\begin{array}{rrl}
{\rm{JD}}_{\rm{max}}^{\rm{phot}} = 2427590 & +~5510 &\times E. \\
\vspace{0.2cm}
          \pm 250 & \pm 90 & \\
\end{array}
\end{equation}
\end{center}

It is worth noticing that both maxima recorded in 1980 and 1994 followed
appreciable brightness decreases which, in turn, could be interpreted
as signatures of an enhanced mass transfer from the cool to the hot
component. A similar effect has been detected in the light curve of the
very slow classical nova V723~Cas (Chochol \& Pribulla, \cite{chp}).

\subsection{$JHK$ data}

\begin{figure}
\begin{center}
\resizebox{8cm}{!}{\includegraphics{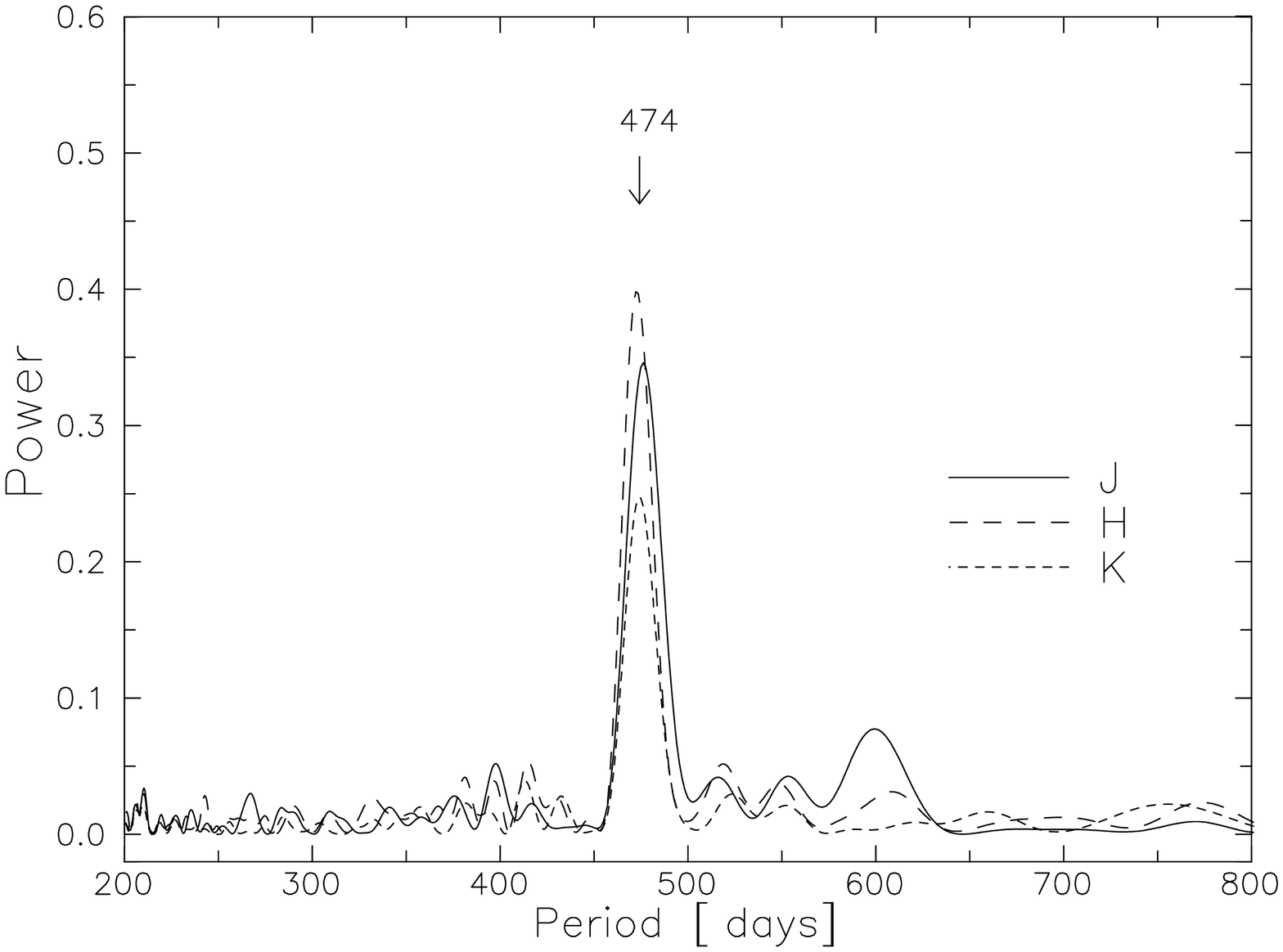}}
\resizebox{8cm}{!}{\includegraphics{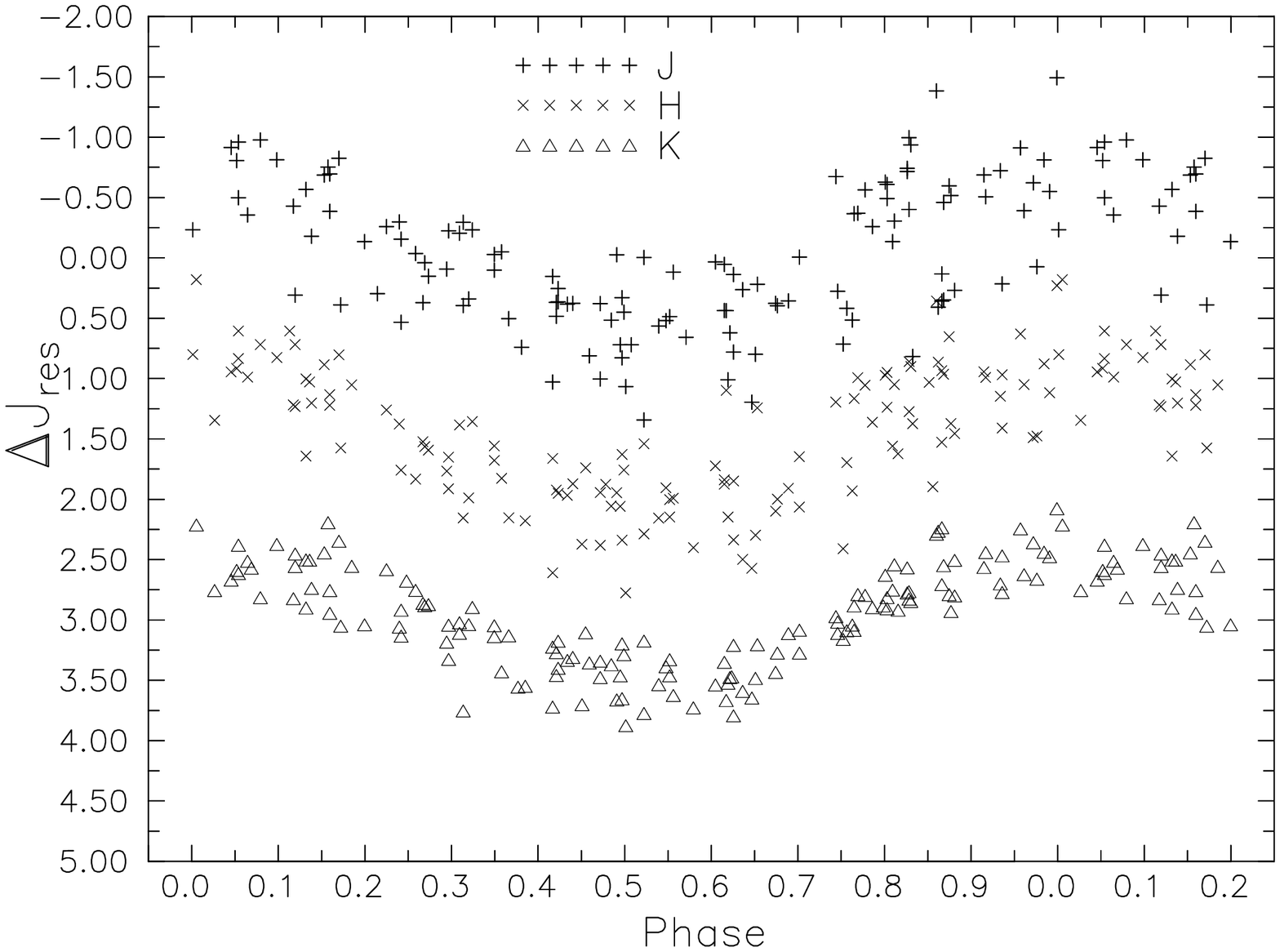}} \caption{Upper panel: the
Fourier power spectra of the pulsation period of the Mira type variable in the
$JHK$ data. Lower panel: the resulting light curves phased with the ephemeris
JD$_{max}$ = 2447442 + 474${\times}E$. The ${\Delta}H_{res}$ and
${\Delta}K_{res}$ are shifted by 1.5 and 3.0 mag, respectively.}
\end{center}
\end{figure}

\begin{figure}
\begin{center}
\hspace*{-0.5cm}
\resizebox{9.5cm}{!}{\includegraphics{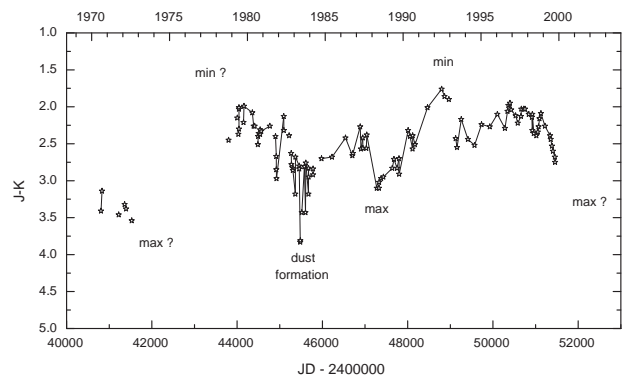}}
\caption{$(J-K)$ color index of V1016 Cyg}
\end{center}
\end{figure}

\begin{figure}
\begin{center}
\resizebox{8cm}{!}{\includegraphics{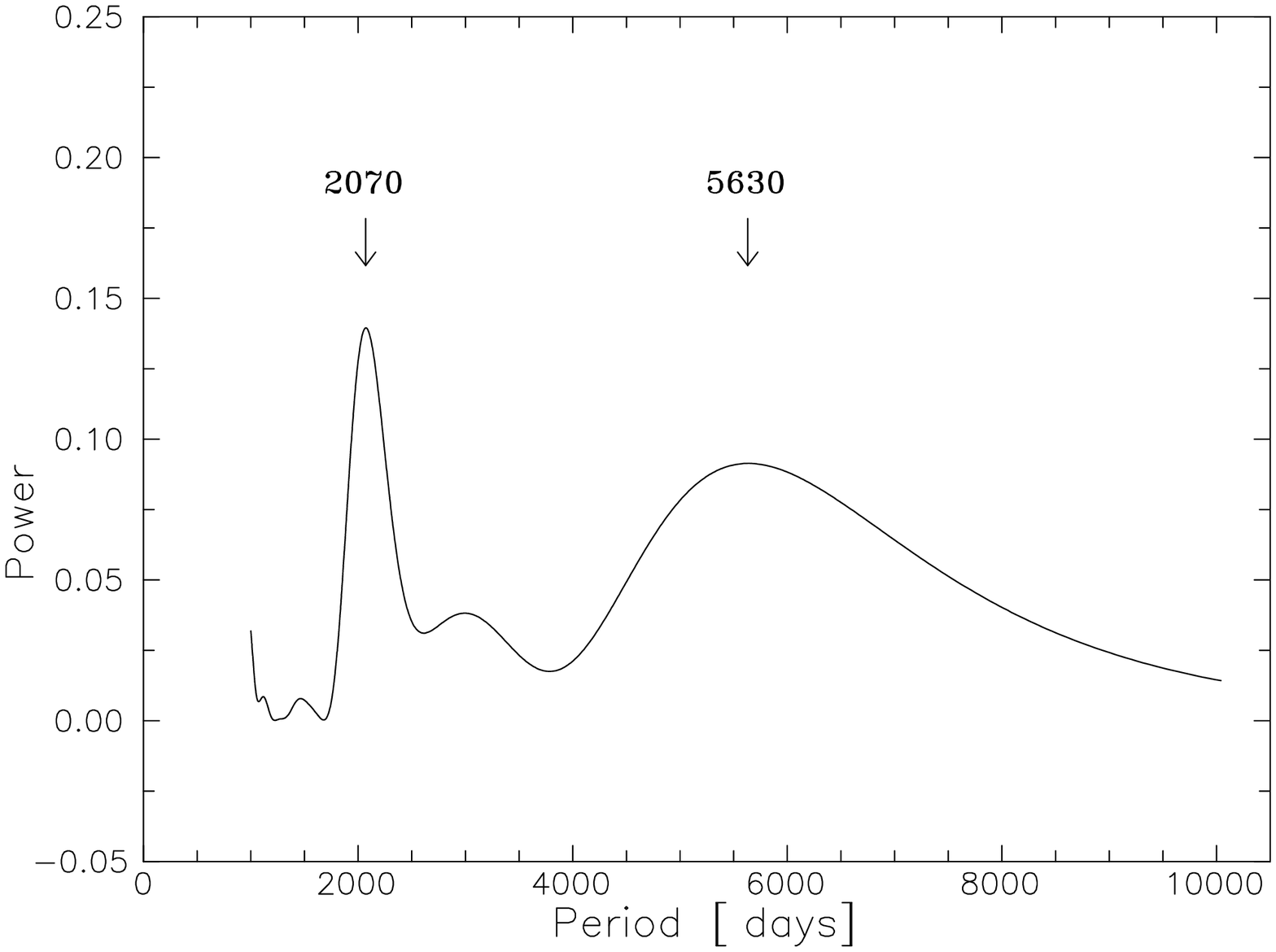}}
\hspace*{-0.6cm}
\resizebox{10.5cm}{!}{\includegraphics{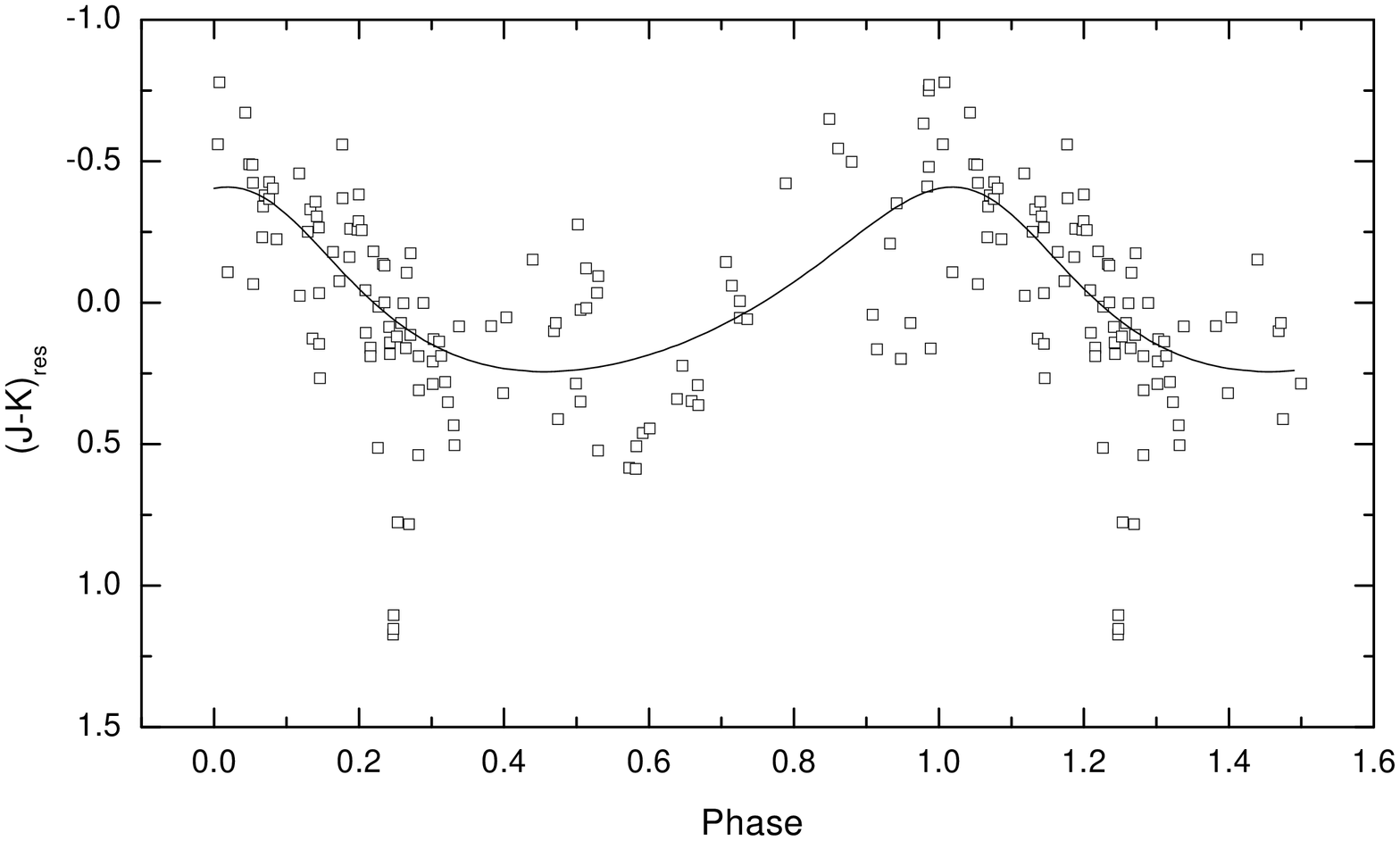}}
\vspace{-2cm}
\caption{Upper panel: the Fourier power spectrum
for the $(J-K)$ residuals in the interval 1000 to 10000 days.
Lower panel: the phase diagram of the above data corresponding to
the ephemeris (1). A trigonometric polynomial has been adopted for the fit.}
\end{center}
\end{figure}

\begin{table}[t]
\begin{center}
\caption{Measured (I) and corrected (II) (with $E(B-V)$=0.28) UV continuum
flux in $10^{-14}$~erg~cm~s$^{-1}$~\AA$^{-1}$}
\footnotesize
\tabcolsep 4pt
\begin{tabular}{cccc|ccc}
\hline
\hline
JD&\multicolumn{3}{c}{1840\AA}&\multicolumn{3}{c}{2530\AA}\\
\hline
2400000+&SWP&I&II&LWR/P&I&II\\
\hline
43752& ~2427    & 15.8&  122.1&  ~2229 &  16.5& 102.6   \\
43920& ~4271    & 15.4&  119.1&  ~3778 &  18.4& 114.0   \\
44049& ~5611    & 16.9&  130.6&        &      &         \\
44072& ~5832    & 17.6&  136.1&  ~5080 &  20.7& 128.3   \\
44076&          &     &       &  ~5136 &  22.6& 140.1   \\
44139&          &     &       &  ~5675 &  19.9& 123.3   \\
44140& ~6617    & 16.3&  125.9&        &      &         \\
44202&          &     &       &  ~6227 &  20.9& 129.5   \\
44268& ~7803    & 13.9&  107.4&        &      &         \\
44474&          &     &       &  ~8593 &  20.8& 128.9   \\
44475& ~9878    & 16.0&  123.7&        &      &         \\
44672& 13431    & 15.0&  115.9&  10095 &  19.1& 118.4   \\
44706&          &     &       &  10344 &  17.8& 110.3   \\
44707& 13704    & 13.8&  106.7&        &      &         \\
44761& 14193    & 12.4&  ~95.8&  10786 &  19.6& 121.5   \\
45065&          &     &       &  12966 &  20.2& 125.2   \\
45192&          &     &       &  13920 &  17.9& 110.9   \\
45193& 17658    & 11.9&  ~91.9&        &      &         \\
45301& 18669    & 14.2&  109.7&  14733 &  15.9& ~98.6   \\
45422& 19566    & 12.4&  ~95.8&  15597 &  14.6& ~90.5   \\
45799&          &     &       &  ~3133 &  12.6& ~78.1   \\
45800& 22701    & 10.0&  ~77.3&        &      &         \\
45822&          &     &       &  ~3261 &  12.8& ~79.3   \\
46045& 24656    & 11.7&  ~90.4&  ~4959 &  12.0& ~74.4   \\
46616&          &     &       &  ~8552 &  12.8& ~79.3   \\
46771&          &     &       &  ~9656 &  11.9& ~73.8   \\
46935& 31006    & 11.1&  ~85.8&  10793 &  12.6& ~78.1   \\
47111& 32296    & 12.6&  ~97.4&  12065 &  13.2& ~81.8   \\
47332& 33783    & 11.4&  ~88.1&  13464 &  12.8& ~79.4   \\
47512& 35047    & 11.8&  ~91.2&  14651 &  12.6& ~78.1   \\
47750&          &     &       &  16106 &  12.9& ~79.9   \\
47751& 36825    & 12.0&  ~92.7&        &      &         \\
47777& 36953    & ~9.7&  ~74.9&  16298 &  14.9& ~92.3   \\
47855&          &     &       &  16822 &  10.6& ~65.7   \\
48004&          &     &       &  17793 &  13.1& ~81.2   \\
48005& 38658    & 10.2&  ~78.8&        &      &         \\
48122& 39486    & 12.0&  ~92.8&  18609 &  12.7& ~78.7   \\
48420&          &     &       &  20582 &  17.3& 107.2   \\
48472& 42161    & 10.9&  ~84.3&  20934 &  12.4& ~76.8   \\
48613&          &     &       &  22055 &  14.9& ~92.4   \\
48614& 43445    & 13.2&  102.1&        &      &         \\
48814&          &     &       &  23483 &  16.0& ~99.2   \\
48917& 46028    & 13.9&  107.4&  24127 &  16.0& ~99.2   \\
49514&          &     &       &  28380 &  15.3& ~94.9   \\
49782& 11301$^a$& 15.7&  121.3&        &      &         \\
49955&          &     &       &  31358 &  14.2& ~88.1   \\
49956& 55707    & 12.0&  ~92.8&        &      &         \\
\hline
\hline
\multicolumn{7}{l}{Note: a - HUT spectrum}
\end{tabular}
\end{center}
\end{table}

Infrared photometry of V1016 Cyg was published by Harvey (\cite{har}),
Kenyon \& Gallagher (\cite{ken}), Lorenzetti et al. (\cite{lor}), Taranova \&
Yudin (\cite{tar2}), Munari (\cite{mun}), Ananth \& Leahy (\cite{anan}),
Kamath \& Ashok (\cite{kam}) and Taranova \& Schenavrin (\cite{tar3}).
We make use here of this whole dataset plus a few unpublished $JHK$
observations obtained in 1993 (kindly provided by F. Strafella) to
newly estimate the period of pulsation of the Mira type variable. The Fourier
period analysis (Fig.~2) applied to the $JHK$ data leads to the values
476$\pm$10, 472$\pm$11 and 474$\pm$10 days for $J$, $H$ and $K$ photometry,
respectively. The adopted period of the Mira type variable pulsations (whose corresponding
phase diagrams are also presented in Fig.~2) is thus $P$ = 474$\pm$6 days.

The long-term behaviour of the $(J-K)$ color index of V1016 Cyg using all
available infrared data is presented in Fig.~3. As pointed out by Whitelock
(\cite{white}), the $(J-K)$ color index in symbiotic Miras is little affected
by the Mira type variable pulsation, being conversely very sensitive to the amount of circumstellar
dust around the cool component. As a consequence, the abrupt change ($\sim$1~mag)
of the IR color recorded in 1983 by Taranova \& Yudin (\cite{tar2}) comes likely from
a short (though strong) dust formation episode. Although orbitally-related dust
obscuration episodes are indeed expected in these systems, we interpret this
unique, major episode which occurred three years after the observed maxima both
in the $U$ passband and space-borne UV (see below) as dust formation in the ejecta
of the symbiotic nova (cf. Bode \cite{bode}).

According to this view, one has to be cautious when interpreting possible
periodicities emerging from the analysis of the long-term behaviour of the $(J-K)$
color index (Fig.~3) which shows an additional maximum in 1988 and a clear
minimum in 1992, together with a more uncertain maximum (in 1973) and
a further possible minimum (in 1977). Indeed, the phenomena lying beneath
the two possible periods (namely, 2070 and 5630 days; see Fig.~4) one can
identify by means of the Fourier analysis of its residuals (after removing
a parabolic trend) is likely quite different. While the longer period,
besides being close to the UBV photometric estimate, could well reflect the
real orbital motion of the system, as suggested by the wave-like variation
exhibited by the $(J-K)$ residuals phased with the ephemeris (1) (see Fig.~4),
the shorter period of 5.6 years, close to the 6-year period interpreted by Munari
(\cite{mun}) as the orbital period of the binary system, is likely due to the
superposition of several effects such as dust formation induced by the activity
of the hot component and possible mass tranfer events from the cool to the hot
component.

\section{UV spectroscopy}

The UV dataset discussed here includes both low and high dispersion IUE
(International Ultraviolet Explorer) spectra spanning the interval from Aug.
1978 to Aug. 1995 (for a complete list see Parimucha (\cite{par}), and a
single HUT (Hopkins Ultraviolet Telescope) spectrum taken on March 6, 1995.
Properly calibrated IUE NEWSIPS data have been retrieved from the IUE
Newly Extracted Spectra (INES) Archive, while the calibrated HUT spectrum
was extracted from the Multimission Archive at STScI (MAST).

Average continuum fluxes in emission-free, 20~\AA\ bins centered
at $\lambda$~1840~\AA\ and $\lambda$~2530~\AA\ have been derived from
low dispersion, large aperture IUE spectra (SWP and LWR/LWP cameras,
respectively), plus the HUT spectrum (for the $\lambda$~1840~\AA\ region
alone). Both measured and dereddened fluxes are given in Table~1
(having adopted $E(B-V)$ = 0.28, according to Nussbaumer \& Schild
\cite{nuss1}). Dereddened continuum fluxes are also shown in Fig.~5.

\begin{figure}
\begin{center}
\resizebox{9.5cm}{!}{\includegraphics{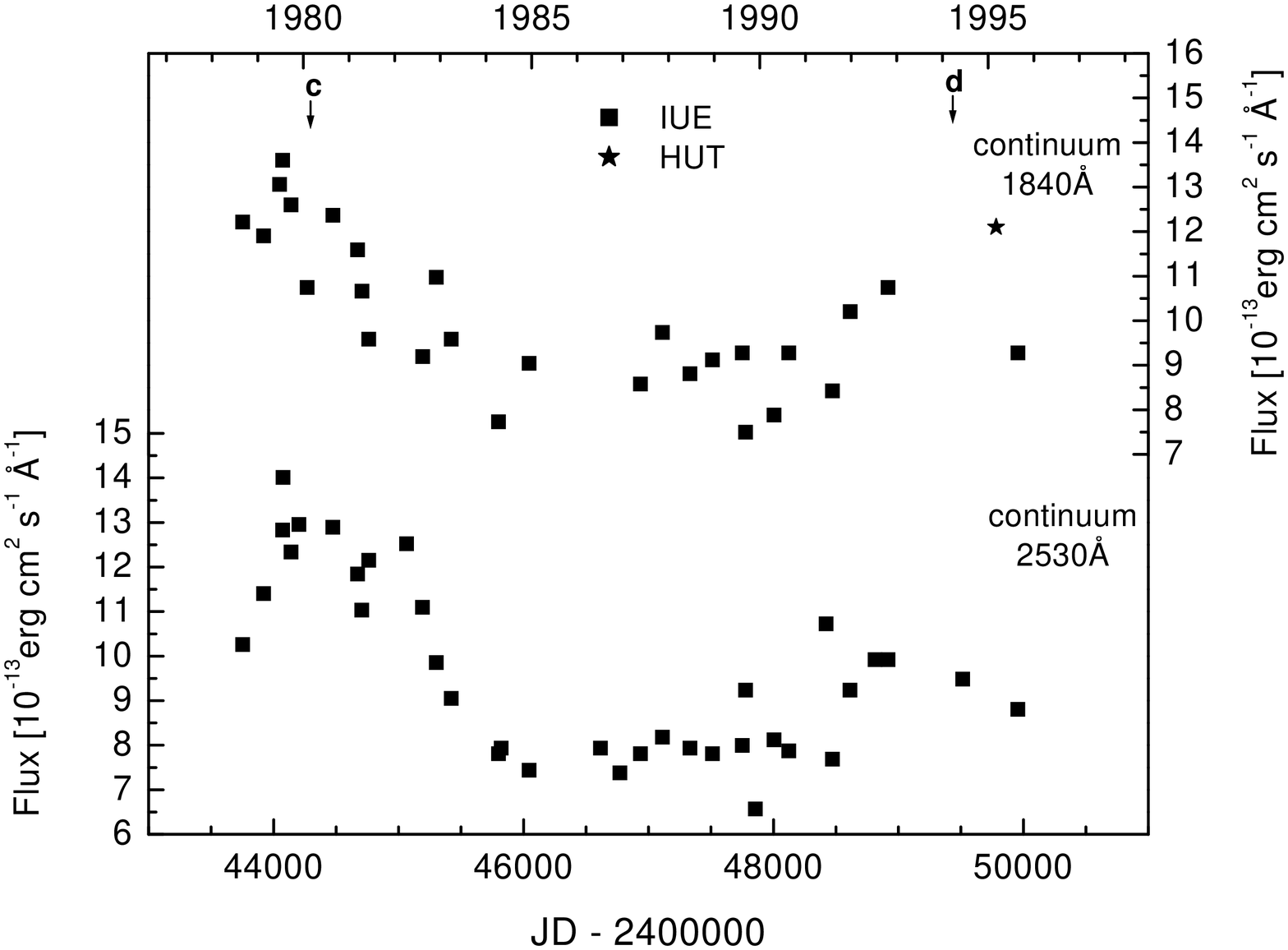}}
\caption{Dereddened continuum fluxes at 1840~\AA\ and 2530~\AA}
\end{center}
\end{figure}

High dispersion, large aperture IUE/SWP spectra were used to derive fluxes
of OIII] ($\lambda\lambda$ 1661,1666~\AA), NIII] ($\lambda$ 1750~\AA),
SiIII] ($\lambda$ 1892~\AA) and CIII] ($\lambda$ 1909~\AA) emission lines.
A single measurement of the OIII] lines flux was obtained also from the
HUT spectrum of 1995. The emission line profiles show a broad component
superposed to the dominant, narrow component (cf. Kindl, Marxer \& Nussbaumer
1982). Fluxes given here are the result of fitting Gaussian profiles to the
latter narrow component. Dereddened fluxes for OIII], NIII] and SiIII]
lines are shown in Fig.~6 and listed (together with the original fluxes)
in Table~2.

\begin{table*}[t]
\begin{center}
\caption{Measured (I) and corrected (II) (with $E(B-V)$=0.28) UV emission line
fluxes in 10$^{-12}$~erg~cm$^{-2}$~s$^{-1}$}
\footnotesize
\begin{tabular}{clcccccccc}
\hline \hline
JD&SWP&\multicolumn{2}{c}{OIII]1661\AA}&\multicolumn{2}{c}{OIII]1666\AA}&\multicolumn{2}{c}{NIII]1750\AA}
&\multicolumn{2}{c}{SiIII]1892\AA}\\
\hline
2400000+&&I&II&I&II&I&II&I&II\\
\hline
44049&  ~5612     &   3.2 &  23.8 &   9.1&  68.0 &   3.7&  29.5 &  9.9 &   80.2 \\
44475&  ~9879     &   3.2 &  23.8 &   8.1&  60.8 &   3.7&  29.5 & 11.0 &   89.1 \\
44672&  13432     &   3.4 &  25.7 &      &       &      &       &      &        \\
44707&  13705     &   3.6 &  27.1 &   9.2&  68.9 &      &       &      &        \\
44761&  14192     &   4.3 &  32.4 &  10.0&  75.1 &      &       &  9.6 &   77.8 \\
45193&  17657     &       &       &      &       &   3.4&  27.1 &  8.3 &   67.2 \\
45301&  18670     &   2.3 &  17.4 &   7.2&  53.8 &   3.3&  26.1 &  8.5 &   68.9 \\
45422&  19568     &   2.1 &  15.5 &   5.5&  41.4 &      &       &      &        \\
45823&  22890     &       &       &      &       &   2.8&  22.6 &  6.4 &   51.8 \\
46045&  24658     &   2.7 &  20.2 &   7.2&  53.5 &   2.7&  21.9 &  5.2 &   42.2 \\
46617&  28618     &   2.2 &  16.2 &   6.5&  48.9 &   2.9&  23.1 &  6.2 &   50.2 \\
46771&  29823     &   1.9 &  14.4 &   5.0&  37.4 &   2.7&  21.3 &  6.2 &   50.2 \\
46935&  31005     &   2.2 &  16.8 &   7.0&  52.5 &   2.5&  20.2 &  6.6 &   53.5 \\
47332&  33785     &   2.2 &  16.5 &   5.6&  41.6 &   2.6&  20.5 &  7.4 &   60.0 \\
47751&  36826     &   1.5 &  11.2 &   4.6&  34.7 &   2.8&  22.4 &  6.8 &   55.1 \\
47777&  36954     &   1.9 &  14.4 &   4.3&  32.0 &   2.4&  19.4 &  5.6 &   45.4 \\
47855&  37673     &   1.9 &  14.4 &   5.3&  39.3 &   2.7&  21.6 &  7.1 &   57.5 \\
48005&  38659     &   1.8 &  13.5 &   4.8&  36.1 &   2.3&  18.7 &  4.9 &   39.7 \\
48420&  41828     &   1.7 &  13.0 &   6.2&  46.4 &   2.5&  20.2 &  6.9 &   55.9 \\
48614&  43446     &   2.5 &  19.0 &   7.0&  52.1 &   2.9&  23.6 &  7.1 &   57.5 \\
48815&  45115     &   2.7 &  19.9 &   7.8&  58.5 &   3.1&  24.6 &  7.6 &   61.6 \\
48917&  46029     &   2.7 &  20.3 &   6.4&  48.1 &   3.3&  26.4 &  7.4 &   59.9 \\
49514&  51059     &   2.4 &  17.6 &   7.5&  56.1 &   3.3&  26.6 &  7.8 &   63.2 \\
49782&  11301$^a$ &   3.4 &  25.9 &   8.8&  66.3 &      &       &      &        \\
49956&  55705     &   2.0 &  14.8 &   5.9&  44.0 &   3.1&  24.6 &  6.7 &   54.3 \\

\hline \hline
\multicolumn{10}{l}{Note: a - HUT spectrum}
\end{tabular}
\end{center}
\end{table*}

For the strong CIII] $\lambda$ 1909~\AA\ which falls by chance in the
overlapping region between the IUE short-wavelength (SWP) and long-wavelength
(LWR/LWP) camera spectral ranges, simultaneous flux estimates obtained
from both cameras are shown in Fig.~7 and listed in Table~3. In order to
make the best use of the whole dataset, Fig.~7 includes
also---as a lower limit---the estimated flux of CIII] in case the line
saturation was marginal (i.e. affecting no more than 2 high-resolution pixels).

It should be stressed that the availability of a {\em large aperture}
spectrum taken on August 31, 1978 (namely, LWR~2228), allowing the measurement
of the CIII] $\lambda$ 1909~\AA\ emission line flux, {\em does assure that the
flux rise exhibited by the UV continua from mid-1978 to mid-1979} (see Fig. 5)
{\em also affects the UV emission lines} so as {\em both} UV continua and emission lines
reflect the 1980 peak of activity of the system recorded in other wavebands
(see above). Strictly speaking, this verification turns out to be not possible
for the remaining set of UV emission lines, as high resolution SWP spectra
assured earlier than June 24, 1979 were obtained through the {\em non-photometric}
IUE small aperture.

One can easily recognize that the evolution of both UV continua
and emission lines from 1978 to 1996 is characterized by two maxima
matching the activity epochs c,d shown by optical photometry.
Moreover, interesting clues to the physical properties of
the system could come from the observed delay of the onset of
emission line activity, whose first maximum (at epoch JD~2444635 $\pm$15)
appears shifted by 410 days from the corresponding epoch for
both UV continua (JD 2444225 $\pm$ 12).

Finally, it should hold reader's interest the fact that the major dust
formation episode discussed in the Sect.~2.2 reflects in a recognizable flux
drop of the majority of UV emission lines in 1983-84 (see Fig.~6).

\begin{figure}
\begin{center}
\resizebox{9cm}{!}{\includegraphics{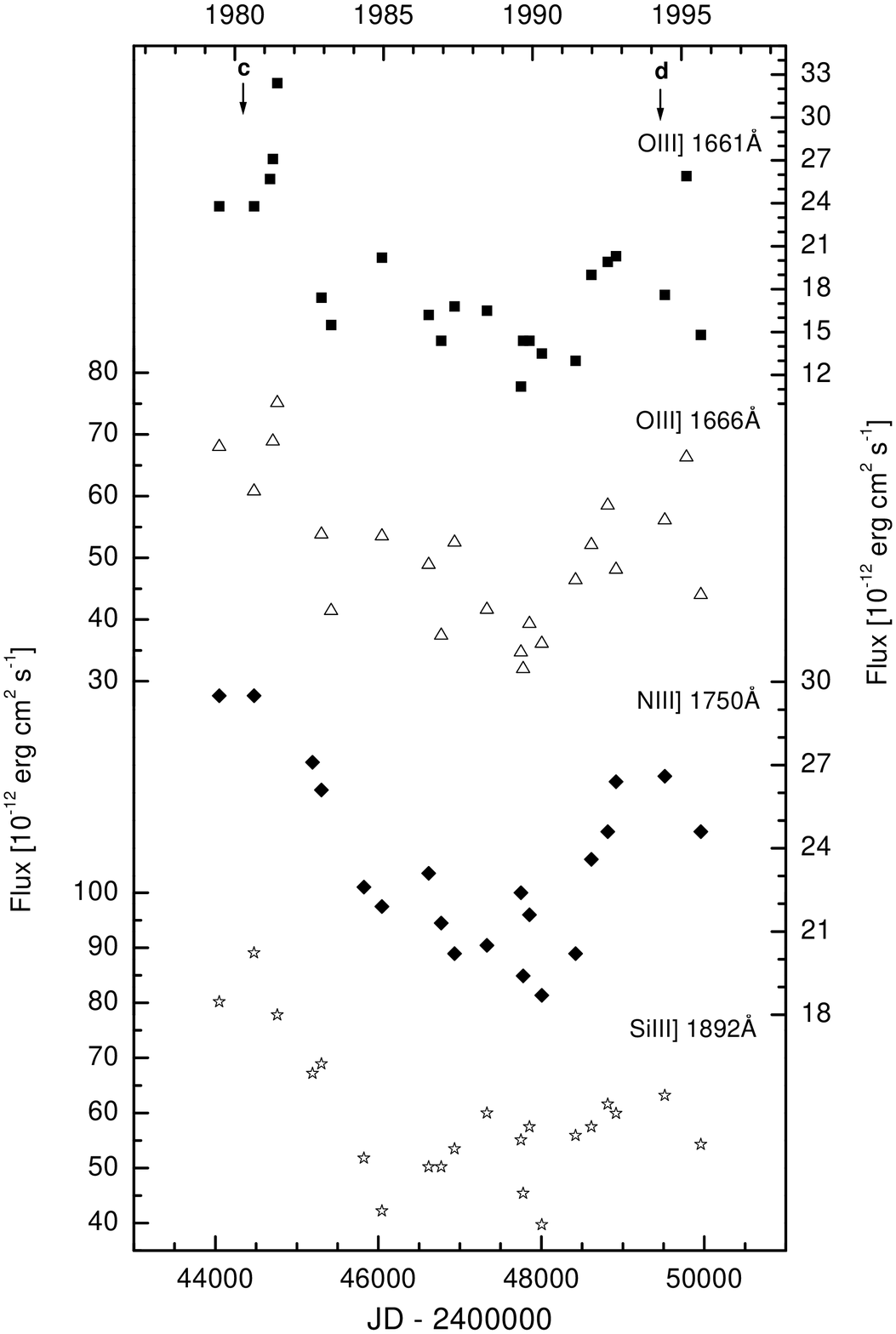}}
\caption{Dereddened UV emission line fluxes}
\end{center}
\end{figure}

\begin{figure}[h]
\begin{center}
\resizebox{9cm}{!}{\includegraphics{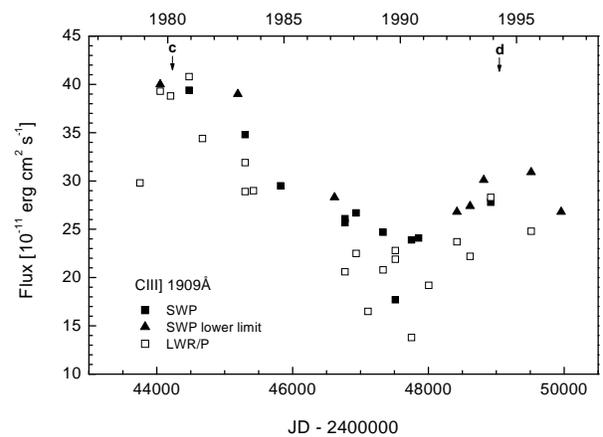}}
\caption{Dereddened fluxes of CIII] $\lambda$ 1909~\AA\ emission line}
\end{center}
\end{figure}

\begin{table}[t]
\begin{center}
\caption{Measured (I) and corrected (II) (with $E(B-V)$=0.28) CIII]1909\AA~emission line
fluxes in 10$^{-11}$~erg~cm$^{-2}$~s$^{-1}$ from SWP and LWP/R spectra}
\footnotesize
\begin{tabular}{clcclcc}
\hline \hline
JD&SWP&I&II&LWR/P&I&II\\
2400000+&&&&&&\\
\hline
43752&            &     &      & ~2228  & 3.6 & 29.8  \\
44049&  5612$^1$  & 4.8 & 40.0 & ~4869  & 4.7 & 39.3  \\
44203&            &     &      & ~6228  & 4.7 & 38.8  \\
44475&  9879      & 4.7 & 39.4 & ~8594  & 4.9 & 40.8  \\
44672&            &     &      & 10096  & 4.1 & 34.4  \\
45193&  17657$^1$ & 4.7 & 39.0 &        &     &       \\
45301&  18670     & 4.2 & 34.8 & 14734  & 3.8 & 31.9  \\
45301&            &     &      & 14735  & 3.5 & 28.9  \\
45422&            &     &      & 15599  & 3.5 & 29.0  \\
45823&  22890     & 3.6 & 29.5 &        &     &       \\
46617&  28618$^1$ & 3.4 & 28.3 &        &     &       \\
46771&  29823     & 3.2 & 26.1 & ~9646  & 2.4 & 19.5  \\
46771&            &     &      & ~9647  & 2.5 & 20.6  \\
46771&  29828     & 3.1 & 25.7 &        &     &       \\
46935&  31005     & 3.2 & 26.7 & 10794  & 2.7 & 22.5  \\
47111&            &     &      & 12066  & 2.0 & 16.5  \\
47332&  33784     & 3.0 & 24.7 & 13463  & 2.5 & 20.8  \\
47512&  35046     & 2.1 & 17.7 & 14652  & 2.7 & 22.8  \\
47512&            &     &      & 14653  & 2.6 & 21.9  \\
47751&  36826     & 2.9 & 23.9 & 16108  & 1.7 & 13.8  \\
47855&  37673     & 2.9 & 24.1 &        &     &       \\
48005&            &     &      & 17795  & 2.3 & 19.2  \\
48420&  41828$^1$ & 3.2 & 26.8 & 20583  & 2.9 & 23.7  \\
48614&  43446$^1$ & 3.3 & 27.4 & 22056  & 2.7 & 22.2  \\
48815&  45115$^1$ & 3.6 & 30.1 &        &     &       \\
48917&  46029     & 3.4 & 27.8 & 24128  & 3.4 & 28.3  \\
49514&  51059$^1$ & 3.7 & 30.9 & 28379  & 3.0 & 24.8  \\
49956&  55705$^1$ & 3.2 & 26.8 &        &     &       \\
\hline \hline
\multicolumn{7}{l}{Note: 1 - lower limit}
\end{tabular}
\end{center}
\end{table}

\section{Discussion}

No doubt the symbiotic nova V1016 Cyg includes a pulsating Mira type variable
and an accreting white dwarf that underwent a thermonuclear outburst
in 1964, leading straight to a nebular spectrum (FitzGerald et al.
\cite{fitz}). According to the ionization model of symbiotic binaries,
the hot luminous component ionizes the neutral wind of the giant giving
rise to the nebula in the system. Such a mechanism is confirmed by
the strict coincidence of the epochs of maxima in the $U$ passband
(observed in 1980 and 1994) and those recorded in the space-UV continuum
combined with the 410-day delay of the OIII], CIII], NIII] and SiIII]
emission line maxima which, in turn, assures that the lines are collisionally
excited in the surrounding nebula when the fast wind from
the hot object interacts with the slow wind from the Mira type variable.

At present, the above well-understood phenomenology leaves
still open two alternative scenarios for V1016 Cygni.

If this symbiotic nova is a wide-orbit binary with a 544-year
period as proposed by Brocksopp et al. (\cite{bbe}), one is
forced to interpret the periodic (15-yr) variations
of the optical and UV continuum as induced by flares of the hot
component triggered by the recurrent enhanced mass loss episodes from
the Mira type variable companion. According to Fleischer et al. (\cite{fgs}),
this mass loss is most pronounced every few periods of the Mira type variable
pulsations. It can trigger individual flares in the accretion disk of
the hot white dwarf. The existence of the disk is supported by the 3D
simulation of the wind accretion by the compact star (Theuns \& Jorissen,
\cite{thjo}).

 Alternatively, V1016 Cygni could be interpreted as a {\em triple} system,
whose {\em inner}, unresolved binary has an {\em orbital} period of 15 years.
If this is the case, flares would be triggered by an enhanced mass transfer
from a cool giant {\em during its periastron passage on the 15-year orbit}.
In this respect, a hint of enhanced mass transfer, coming just before
the weak flare in 1994, can be found in the long-term infrared photometry
performed by Taranova \& Schenavrin (\cite{tar3}) showing a brightness increase
in the $J$ and $H$ passbands in 1992. Moreover, the coincidence of the
observed maximum of the $J-K$ index in 1988 (Fig.~3) and the wide minima
of the UV continua (Fig.~5) would constrain to that epoch
the inferior conjuction of a cool giant along its 15-year orbit.

\begin{acknowledgements}

This work was accomplished as a part of a PhD thesis of \v{S}.P. and has been
supported by VEGA Grant 2/1157 of the Slovak Academy of Sciences. The
authors would like to thank to F. Strafella for providing his
unpublished IR observations and A. Skopal for critical reading of
the manuscript.

\end{acknowledgements}

\end{document}